\documentclass[reprint,amsmath,amssymb,aps,prb,longbibliography,superscriptaddress]{revtex4-1} 
\usepackage{graphicx}
\usepackage[space]{grffile}
\usepackage{latexsym}
\usepackage{textcomp}
\usepackage{longtable}
\usepackage{multirow,booktabs}
\usepackage{amsfonts,amsmath,amssymb}
\usepackage{natbib}
\usepackage{url}
\usepackage{hyperref}
\usepackage{color}
\usepackage{datetime}
\usepackage[dvipsnames]{xcolor}
\usepackage[normalem]{ulem}

\hypersetup{colorlinks=false,pdfborder={0 0 0}}
\newif\iflatexml\latexmlfalse
\usepackage[utf8]{inputenc}
\usepackage[ngerman,english]{babel}
\usepackage{babel}

\newcommand{\bra}[1]{\langle\,#1\,|\,}
\newcommand{\ket}[1]{\,|\,#1\,\rangle}

\newcommand{\braket}[2]{\left\langle\,#1\,|\,#2\,\right\rangle}
\def\rv{\mathbf{r}}

\def\ra{\mathbf{\hat{r}}}
\def\qa{\mathbf{\hat{q}}}
\def\kv{\mathbf{k}}
\def\Kv{\mathbf{K}}
\def\pv{\mathbf{p}}
\def\qv{\mathbf{q}}
\def\Gv{\mathbf{G}}
\def\pv{\mathbf{p}}

\def\NA{N_\xi}
\def\NO{N_\zeta}
\def\qr{\mathbf{q}\!\cdot\!\rv}
\def\QR{\mathbf{q}\cdot\rv}
\def\iaa{\AA$^{-1}$}
\def\qIOO{$\mathbf{q}\!\parallel\![100]$}
\def\qIIO{$\mathbf{q}\!\parallel\![110]$}
\def\qIII{$\mathbf{q}\!\parallel\![111]$}

\begin{document}

\title{An all-electron product-basis set: application to plasmon anisotropy in simple metals}
\author{J. A. Budagosky}
\affiliation{Donostia International Physics Center (DIPC), Paseo Manuel de Lardizabal 4,
  20018 Donostia/San Sebasti\'{a}n, Basque Country,
  Spain}

\author{E. E. Krasovskii}
\affiliation{Donostia International Physics Center (DIPC), Paseo Manuel de Lardizabal 4,
  20018 Donostia/San Sebasti\'{a}n, Basque Country,
  Spain}
\affiliation{Departamento de F\'{i}sica de Materiales, Facultad de Ciencias Qu\'{i}imicas, Universidad del Pais Vasco/Euskal Herriko Unibertsitatea, Apdo. 1072, 20080 Donostia/San Sebasti\'{a}n, Basque Country, Spain}
\affiliation{IKERBASQUE, Basque Foundation for Science, 48013 Bilbao, Basque Country, Spain}


\begin{abstract}
  An efficient basis set for products of all-electron wave functions is proposed, which
  comprises plane waves defined over the entire unit cell and orbitals confined to small
  non-overlapping spheres. The size of the set and the basis functions are, in principle,
  independent of the computational parameters of the band structure method. The approach
  is implemented in the extended LAPW method, and its properties and accuracy are discussed.
  The method is applied to analyze the dielectric response of the simple metals Al, Na, Li,
  K, Rb, and Cs with a focus on the origin of the anisotropy of the plasmon dispersion in Al
  and Na. The anisotropy is traced to tiny structures of the one-particle excitation spectra
  of Al and Na, and relevant experimental observations are explained.
\end{abstract}

\maketitle

\section{Introduction}
\label{introduction}
The microscopic dielectric function (DF) $\epsilon(\rv,\rv';\omega)$ is a key ingredient in
the theory of the ground state~\cite{Giuliani_2005},
quasiparticles~\cite{Hedin_1965,Aryasetiawan_1998}, optical~\cite{Haug_2004}
and plasmonic~\cite{Pines_1964,Feibelman_1982,Liebsch_1997,Pitarke_2007} excitations as well as
in the electron spectroscopies where the microscopic electric field in the solid is a
crucial aspect: photoemission at low photon energies~\cite{Krasovskii_2010}, laser-assisted
time-resolved spectroscopy~\cite{Siek_2017}, or the theory of energy losses by quantum
particles~\cite{Nazarov_2017}. The variety of applications and the growing demand for a
detailed description of the DF calls for the development of an efficient basis set to express
the relevant operators in order to facilitate {\it ab initio} calculations of the DF.

A general and rigorous analysis of the basis set problem was presented
by Harriman~\cite{Harriman1986}, and various practical schemes have been implemented. The
simplest case are pseudopotential methods~\cite{Abinit_1992,Bloechl_1994}, where the plane-wave
(PW) basis for the Bloch wave functions $\psi^{\kv}_{\lambda}$ is ideally suited for the Fourier
representation of the dielectric matrix $\epsilon_{\Gv\Gv'}(\qv,\omega)$, which immediately
follows from the Fourier expansion of the products $\psi^{\kv *}_{\lambda'}\psi^{\kv-\qv}_{\lambda}$.
Furthermore, the accuracy of both $\psi^{\kv}_{\lambda}$ and $\epsilon(\rv,\rv';\omega)$ is
consistently controlled by a cutoff in the reciprocal space. Another obvious choice is
a real-space grid~\cite{Brodersen2002,Mortensen2005}, however, the experience with the
projected augmented wave method shows that atomic-orbital basis is computationally more
efficient~\cite{Yan2011}. Typical implementation of the orbital basis in the context of
pseudopotentials and accompanying approximations are discussed in
Refs.~\onlinecite{Blase2004,Umari2009}.
The problem becomes nontrivial in all-electron methods, where the basis functions have
complicated shape, and their products are unwieldy. In addition, the resulting product
set is non-orthogonal and overcomplete, and there is no {\it a priori} criterion to
reduce the set and to control its convergence. For orbital basis sets, methods of
numerical elimination of redundant products were suggested by Aryasetiawan and
Gunnarsson~\cite{AryasetiawanGunnarsson_1994} for muffin-tin orbitals and by
Foerster~\cite{Foerster2008} for atomic orbitals. The most accurate wave functions are
provided by the augmented plane wave (APW) formalism, where the basis functions are still more
complicated~\cite{Slater_1937,Andersen_1975,Koelling_1975,Singh_1991,Krasovskii_1997}. In the
LAPW-based (linear APW) codes~\cite{Jiang_2013,Friedrich2010,Nabok2016}, criteria similar to
those of Refs.~\onlinecite{AryasetiawanGunnarsson_1994,Foerster2008} are applied to obtain a
reduced set from the products of APWs. However, no attempt has been made to construct a
universal basis to parametrize the products.

Here, we propose a basis set to calculate the $\epsilon$ matrix out of all-electron wave
functions, which is suitable for (but not limited to) the APW representation. The product
basis set consists of plane waves that are defined throughout the unit cell (not just in the
interstitial region, in contrast to
Refs.~\onlinecite{Jiang_2013,KOTANI2002,Friedrich2010,Friedrich2011,Nabok2016}) and
orbitals centered at atomic sites and confined to small non-overlapping spheres. We refer to
the latter as island orbitals (IOs) to distinguish them from the localized orbitals (LOs) of
the extended LAPW methods~\cite{Singh_1991,Krasovskii_1997}. The basis
functions are derived from an approximation to the all-electron wave functions, which also
has the PW+IO structure, and whose accuracy is regulated by the angular-momentum cutoff of
the orbital part and by the $|\kv+\Gv|$ cutoff of the plane wave part.
Owing to the orthogonality of the plane waves and to the finite domain of the orbitals,
the basis set provides an efficient scheme for the matrix elements
$\bra{\kv+\qv\lambda'}\exp[i(\qv+\Gv)\cdot\rv]\ket{\kv\lambda}$. Thus, the Fourier
representation $\epsilon_{\Gv\Gv'}(\qv,\omega)$ is readily obtained, which is convenient, in
particular because the Coulomb interaction becomes diagonal and because the reciprocal lattice
vector $\Gv$ is a natural cutoff parameter to systematically refine the accuracy of the DF.

To demonstrate the viability of the method we address the $\qv$-dependent longitudinal DF
of simple metals, focusing on the anisotropy of the plasmon dispersion. We analyze the
similarities and differences in the anisotropy scenarios in Al and Na and explain the
experimentally long-known opposite behavior of the anisotropy of the Drude plasmon relative
to the low-energy zone boundary collective state (ZBCS)~\cite{Foo_1968} in the two metals.
Further, we compare Al and Na with the less free-electron-like alkali metals Li, K, Rb, and Cs.

The paper is organized as follows. In the next section we describe the PW+IO representation of
the wave functions and the resulting formalism for the DF. The accuracy and convergence of the
product basis are analyzed in Appendix. Computational aspects not related to the new method are
presented in Sec.~\ref{computation}. In Sec.~\ref{PD}, plasmon dispersion in aluminum and alkali
metals is discussed.

\section{formalism}
\label{theoretical_method}
Of all the band structure methods, the formalism of augmented plane waves introduced by
Slater~\cite{Slater_1937} has the least limitations regarding the accuracy of the wave
functions. Here, we consider an implementation for a basis of energy-independent
APWs~\cite{Andersen_1975,Koelling_1975} extended by localized
orbitals~\cite{Singh_1991,Krasovskii_1997}.

\subsection{Filtering the all-electron wave function}\label{filter}
The wave function $\psi_{\lambda}^{\kv}$ for the Bloch vector $\kv$ and band number $\lambda$
is a sum of $\NA$ APWs $\xi(\rv)$ and $\NO$ localized orbitals $\zeta(\rv)$. For one atom per
unit cell located at $\rv=0$ it reads
\begin{equation}\label{eq-ELAPWWF}
\psi_{\lambda}^{\kv}(\rv) = \sum_{\Gv}^{\NA}p_{\Gv}^{\lambda\kv}\xi_{\Gv}^{\kv}(\rv) +
\sum_{lm}\sum_{\nu}^{\bar\nu_l}
q_{\nu lm}^{\lambda\kv} \zeta_{l\nu}(r) Y_{lm}(\ra).
\end{equation}
Here $p$ and $q$ comprise the set of variational coefficients, $p$ being coefficients of the
APWs and $q$ of the LOs, and $\Gv$ are reciprocal lattice vectors. The APW $\xi_{\Gv}^{\kv}$ is
smoothly continuous everywhere in the unit cell, and outside the muffin-tin (MT) spheres it
coincides with the plane wave $\exp[i(\kv+\Gv)\cdot\rv]$. The LOs vanish with their radial
derivatives at the muffin-tin sphere and remain zero in the interstitial region. Subscript
$\nu$ indicates the radial part of the local orbital, so $\NO=\sum_l\bar\nu_l(2l+1)$.
\begin{figure}[b] 
\includegraphics[trim=0.cm 0.cm 0cm 8cm, clip=true,width=0.46\textwidth]{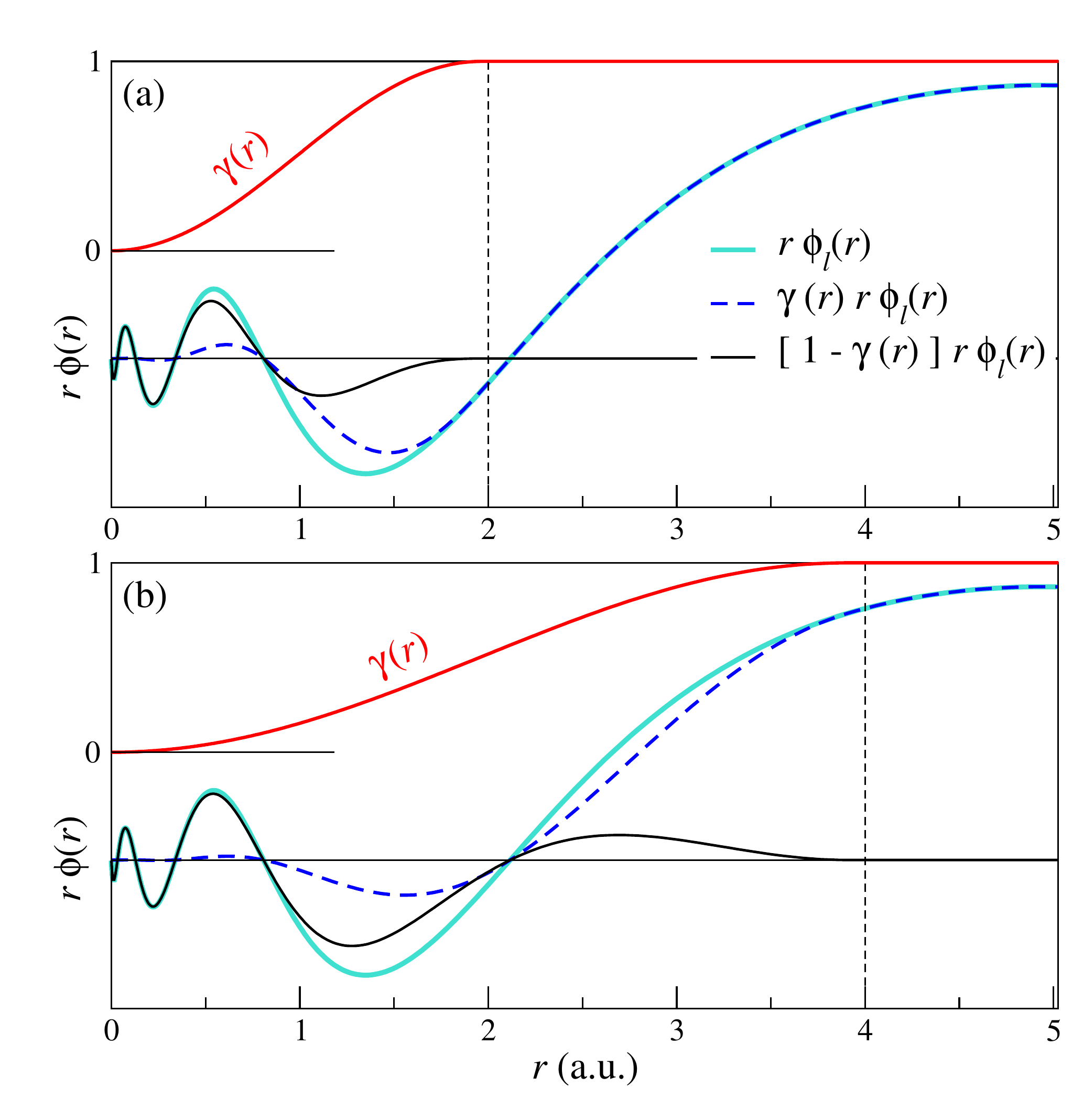}
\caption{$\gamma$-partitioning in the sphere of Cs for $S_\gamma=2$~a.u.~(a)
    and $S_\gamma=4$~a.u.~(b): radial solution (thick light-blue line)
    is multiplied by $\gamma(r)$ (thick red line). The gouging radii $S_\gamma$ are indicated
    by vertical dashed lines, MT radius is 5.02~a.u. The radial function is for
    $l=0$, and energy is at the Fermi level.}
\label{fig-gouging}
\end{figure} 

In the sphere the angular-momentum expansion of the APW of a wave vector $\Kv=\kv+\Gv$ reads
\begin{equation}
\label{eq-APWYLM}
\xi_{\Gv}^{\kv}(\rv)=
\sum_{lm}[\,A_{lm}(\Kv)\phi_{l}(r)+B_{lm}(\Kv)\dot{\phi}_{l}(r)\,]\,
Y_{lm}(\mathbf{\ra}),
\end{equation}
where $\phi_l$ is a solution of the radial Schr\"odinger equation and $\dot\phi_l$ is its
energy derivative~\cite{Andersen_1975,Koelling_1975}. Coefficients $A_{lm}$ and $B_{lm}$ are
determined from the condition that the APW be smoothly continuous at the sphere boundary,
$r=S$. Thus, for each $l$ the radial basis comprises $\bar n_l=\bar\nu_l+2$ functions $u_{ln}$,
where $u_{l1}=\phi_l$, $u_{l2}=\dot\phi_l$, and the radial part of the LO is a linear combination
of three functions: $\zeta_{l\nu}=u_{l(\nu+2)} +a_{l\nu}\phi_l +b_{l\nu}\dot\phi_l$.
Usually, $u_{l(\nu+2)}$ are also radial solutions for different energies, although in some
applications they may be more complicated functions~\cite{Krasovskii_2001,MICHALICEK2013}.

A straightforwardly constructed set of products of $N=\NA+\NO$ functions would comprise
$N(N+1)/2$ terms, of which a much smaller number are linearly independent and physically
relevant. There is no universal recipe to {\it a priori} select the optimal subset---without
reference to the specific shape of the APWs---although some intuitive criteria and practical
schemes were suggested in Refs.~\onlinecite{AryasetiawanGunnarsson_1994,Jiang_2013}. Here we
develop an approach that avoids an explicit construction of the products of the APWs, but
instead employs an approximate (filtered) representation of the wave functions to generate
the product basis. This approximate representation has the property that both the accuracy
of the filtered wave functions and the completeness of the product basis are naturally
controlled by a {\em spatial resolution criterion}, i.e., by the $\Gv$-vector cutoff of the PW
set and by the $lm$-cutoff of the IO set.

To arrive at the optimal partitioning between the PW and IO components of the
filtered wave function, we first modify the original wave function $\psi$ by
damping its rapid oscillations in the vicinity of the nuclei: within a sphere of
radius $S_\gamma$ (smaller than the muffin-tin radius, see Fig.~\ref{fig-gouging})
we multiply $\psi$ by a function $\gamma(r)$ that is zero at $r=0$ and steadily
grows to reach unity at $r=S_\gamma$:
\begin{eqnarray}\label{eq-gougingfunc}
\gamma(r)=\left\{%
\begin{array}{ll}
  \dfrac{1}{2}\left(1-\cos\dfrac{\pi r}{S_\gamma}\right)& \text{for}~ r \leq S_\gamma, \\
  1  & \text{for}~ r > S_\gamma. \\
\end{array}%
\right.
\end{eqnarray}
The gouged function $\gamma\psi$ has a rapidly convergent plane-wave
expansion~\cite{Krasovskii_1999a,Krasovskii_1999}, which constitutes
the Fourier part $\psi^\text{F}$ of the filtered wave function:
\begin{equation}\label{eq-FI}
\psi = \gamma\psi + (1-\gamma)\psi \approx \psi^\text{F} + (1-\gamma)\psi
                                   \approx \psi^\text{F} + \psi^\text{I}.
\end{equation}
Here $\psi^\text{I}$ is an approximation to $(1-\gamma)\psi$ obtained
by truncating the angular-momentum expansion of $\psi$. It consists
of isolated islands around the nuclei, and the smaller the radius $S_\gamma$
of the island the faster converges the $lm$ series of $\psi^\text{I}$
and the slower does the PW series of $\psi^\text{F}$. For sufficiently small
$\gamma$~spheres, $\psi^\text{I}$ can be completely neglected to a good
approximation~\cite{Krasovskii_1999,Krasovskii_1999a}. In Appendix,
we present a detailed study of the convergence and accuracy of the
$\psi^\text{F} + \psi^\text{I}$ representation.

\subsection{Density matrix elements}\label{dmme}
Let us consider the operator $\hat{o}=\exp(i\qr)$. Using the notation
$\psi_\lambda=\braket{\rv}{\psi_\lambda}$, $\psi^\text{F}_\lambda=\braket{\rv}{F_\lambda}$, and
$\psi^\text{I}_\lambda=\braket{\rv}{I_\lambda}$ for the approximate wave functions
$\psi_\lambda=\psi^\text{F}_\lambda+\psi^\text{I}_\lambda$ given by Eq.~(\ref{eq-FI}) we write
the matrix element $\bra{\psi_\lambda}\hat{o}\ket{\psi_\mu}$ as a sum
of an integral over the entire unit cell $\bra{F_\lambda}\hat{o}\ket{F_\mu}$ and
three integrals over the small $\gamma$-spheres:
\begin{eqnarray}\label{eq-integrals}
\bra{\psi_\lambda}\hat{o}\ket{\psi_\mu}&=&\bra{F_\lambda}\hat{o}\ket{F_\mu}\\
&+&\bra{F_\lambda}\hat{o}\ket{I_\mu}+\bra{I_\lambda}\hat{o}\ket{F_\mu} \nonumber
+\bra{I_\lambda}\hat{o}\ket{I_\mu}.\quad
\end{eqnarray}
The first term in the right hand side is readily calculated via plane waves, and the integrals
over the $\gamma$-spheres can be calculated in the angular-momentum representation in
view of the relation $\psi^\text{F}\approx\gamma\psi$, see Eq.~(\ref{eq-FI}). Their sum reduces
to $\bra{\psi_\lambda}\hat{o}\ket{\psi_\mu}_\gamma - \bra{F_\lambda}\hat{o}\ket{F_\mu}_\gamma$, where
the subscript $\gamma$ indicates that the integration is limited to the $\gamma$-spheres. The
computational efficiency of this scheme stems from the following properties: First, the $lm$
series converges fast because the $\gamma$-spheres are small, and it is helpful that the
coefficients are the same for the wave function $\psi$ and for its Fourier-filtered part
$\psi^\text{F}$. Second, the plane-wave expansion of $\gamma\psi$ contains a reasonable number
of plane waves because---in contrast to the energy-eigenvalue problem---a high accuracy close
to the nucleus is not needed (see Appendix). Note that the approximate function~(\ref{eq-FI})
is smoothly continuous by construction at any $l$-cutoff, whereas in the original APW
representation one has to include rather high angular momenta to achieve the continuity.
This property is important, in particular for the construction of the effective potentials
from orbital-dependent functionals~\cite{Betzinger2011} .

\begin{figure}[t] 
\includegraphics[trim=1.1cm 1.2cm 0cm 14.0cm, clip=true,width=0.495\textwidth]{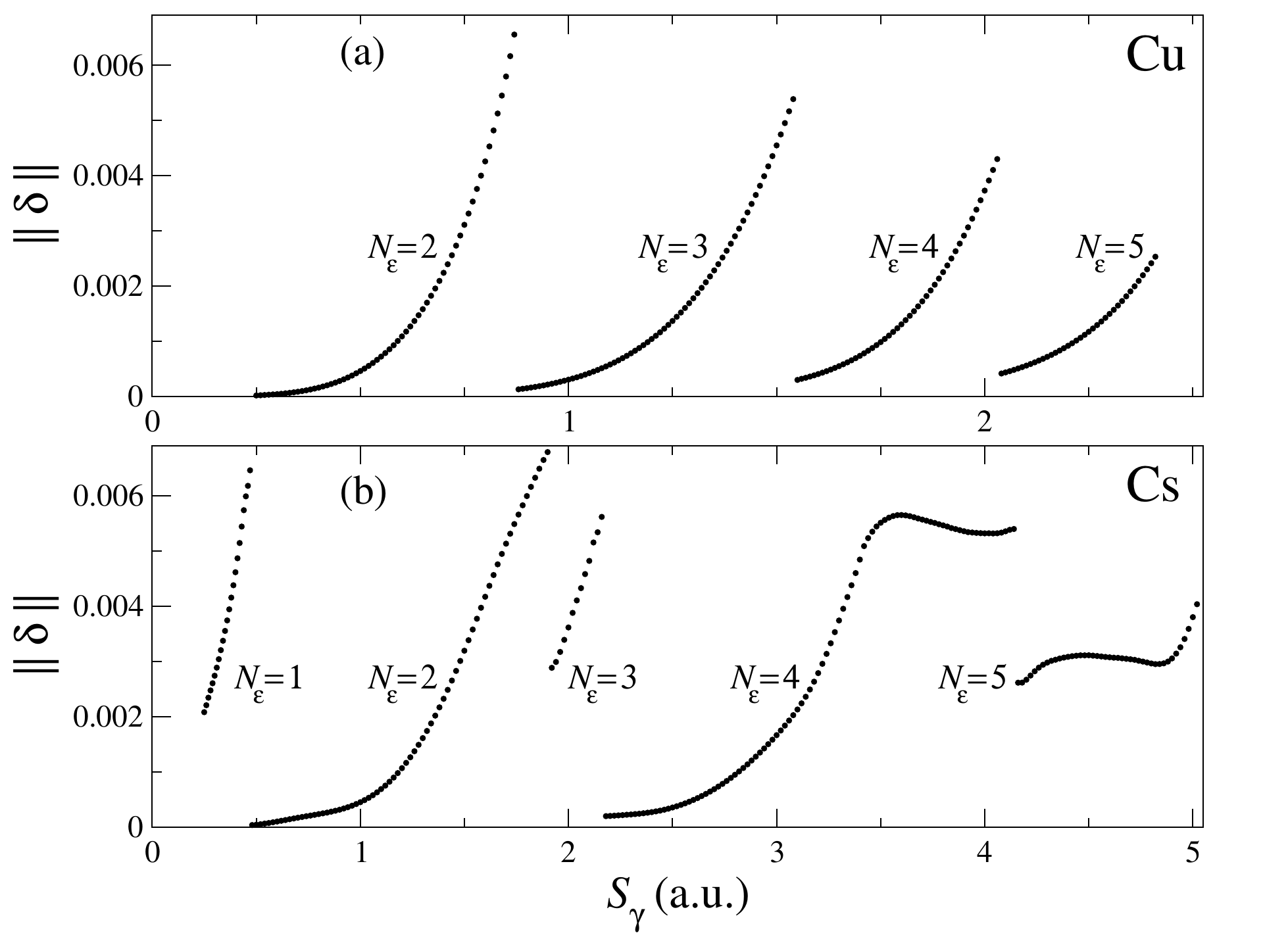}
\caption{Accuracy of the representation of the radial density matrix $U_{ll'nn'}(r)$ by a
reduced orthogonal basis set, see the last paragraph of Sec.~\ref{dmme}. The error
$\|\,\delta(N_\varepsilon,S_\gamma)\,\|$ is shown for the products of the $d$-orbitals of Cu~(a)
and $p$-orbitals of Cs~(b). The overlap eigenvalue cutoff is $\varepsilon=0.0001$.
}
\label{selection}
\end{figure}      
The operator $\exp(i\QR)$ is diagonal in the PW basis, so
the first term in Eq.~(\ref{eq-integrals}) is easy to calculate. The contribution from
the islands is obtained from the angular-momentum decomposition of the wave functions
inside the muffin-tin spheres
\begin{equation}\label{eq:amd}
\psi_{\lambda}^{\kv}(\rv)=\sum_{l}^{l_{\text{max}}}\sum_{n}^{\bar n_l}\sum_{m=-l}^{l}
C_{lnm}^{\kv\lambda}u_{ln}(r)Y_{lm}(\ra)
\end{equation}
using the Rayleigh expansion of $\exp(i\qr)$:
\begin{widetext}
\begin{equation}
\label{eq-R}
\bra{\psi_{\lambda'}^{\kv+\qv}} e^{i\qv\cdot\rv}
\ket{\psi_\lambda^\kv}_\gamma -
\bra{F_{\lambda'}^{\kv+\qv}} e^{i\qv\cdot\rv}
\ket{F_\lambda^\kv}_\gamma =
\sum_{ll'}^{l_{\text{max}}}\sum_{nn'}\sum_{mm'}
\left[C^{\kv+\qv\lambda'}_{l'n'm'}\right]^*
C^{\kv\lambda}_{lnm}\sum_{l''}^{l''_{\text{max}}}T_{ll'mm'}^{l''}(\qa)
\int\limits_0^{S_\gamma}\!j_{l''}(qr)U_{ll'nn'}(r)\,dr,
\end{equation}
\end{widetext}
where the angular integration yields
\begin{equation}\label{eq-Y}
T_{ll'mm'}^{l''}(\qa)=
{4\pi}i^{l''}Y^*_{l''m''}(\mathbf{\hat{q}})\int Y_{l'm'}^*Y_{l''m''}Y_{lm}\;d\mathbf{\ra}.
\end{equation}
Here $lm$ and $l'm'$ refer to the angular-momentum decomposition of $\psi_{\lambda}^{\kv}$ and
$\psi_{\lambda'}^{\kv+\mathbf{q}}$, respectively, and $l''$ to the Rayleigh expansion of
$\exp(i\qr)$.
The radial functions
\begin{equation}\label{eq-U}
U_{ll'nn'}(r)=u_{l'n'}(r)u_{ln}(r)[1-\gamma(r)^2]r^2
\end{equation}
are products of the radial parts of APWs multiplied by the confining function
$[1-\gamma(r)^2]r^2$. Because the radii $S_\gamma$ are independent of the other computational
parameters they can be chosen rather small, typically 1 to 2~a.u., so that the angular-momentum
sums can be truncated at rather low $l$, as will be demonstrated in the next section. To
summarize, the PW+IO product basis set consists of plane waves $\exp[i(\qv+\Gv)\cdot\rv]$ and
island orbitals $U_{ln}(r)Y_{lm}(\ra)$ whose radial shape is given by Eq.~\eqref{eq-U} and
angular part by Eq.~\eqref{eq-Y}. Its size scales linearly with the size of the unit cell,
and the IO part can be further reduced by removing the linearly dependent
IOs~\cite{AryasetiawanGunnarsson_1994,Foerster2008,Friedrich2010,Jiang_2013,Koval2014}.

According to Eq.~\eqref{eq-U}, for each pair of $l$ and $l'$ the set of radial products
comprises $\bar n_l\times\bar n_{l'}$ functions, and large $\bar n_l$ may be needed to describe
a wide energy interval. For example, in order to achieve convergence of the $GW$ method with
respect to the number of unoccupied states it may be necessary to include up to $\bar n_l=8$
radial functions per $lm$-channel~\cite{Friedrich2011,Nabok2016,Jiang2018}. A straightforward
inclusion of all the products $u_{l'n'}u_{ln}$ may lead to an excessively large and linearly
dependent basis set. Here, we can take advantage of the fact that $U_{ll'nn'}(r)$ are restricted
to a close vicinity of the potential singularity, where the radial solutions change very slowly
with energy~\cite{Andersen_1975,MICHALICEK2013}, so the number of physically relevant $u_{ln}$
is much smaller. Moreover, their shape is determined by the potential singularity, and it is
practically independent of the crystal potential. Thus, for a given $S_\gamma$, the functions
$U_{ll'nn'}(r)$ can be tabulated for each element.

To demonstrate how a relevant set can be selected out of a full set of functions $U_{ll'nn'}(r)$,
let us take the MT sphere of Cu as an example and consider the contribution of the $d$-orbitals
to the spherical part $l''=0$ of Eq.~\eqref{eq-R}. The 3$d$ band is described by the
radial solution at the energy $E=-3$~eV relative to the Fermi level and by its energy
derivative. The radial set is extended by a 4$d$ and a 5$d$ function at $E=40$ and 133~eV,
respectively. This gives rise to $N_{U}=10$ functions $U_{llnn'}(r)$ with $l=2$. We now
diagonalize the overlap matrix of the functions $U_{llnn'}(r)$ and retain only the eigenvectors
of eigenvalues larger than some predefined $\varepsilon$. The number $N_\varepsilon$ of the
retained functions is the smaller the smaller the gouging radius $S_\gamma$. We then fit
the $N_U$ original functions with the $N_\varepsilon$ orthogonal functions and in
Fig.~\ref{selection}(a) present the maximal error $\|\,\delta(N_\varepsilon,S_\gamma)\,\|$
as a function of $S_\gamma$. Figure~\ref{selection}(b) shows the performance of the
orthogonal set in the sphere of Cs for four $p$-orbitals
comprising $\phi$ and $\dot\phi$ at the 5$p$ branch and two $\phi$ functions at the 6$p$ and
7$p$ branches. Thus, the accuracy with which the radial part of the product set is represented
is flexibly adjusted by the gouging radius and by the number of the orthogonal basis functions.

The data on the accuracy of the product-set expansion regarding the number of PWs in
$\psi^\text{F}$ and IOs in $\psi^\text{I}$, Eq.~\eqref{eq-FI}, as well as on the $l''$
convergence of the $\exp(i\qr)$ operator, Eq.~\eqref{eq-R}, are given in Appendix. For
sufficiently large $S_\gamma$ the number of PWs may be chosen comparable to the number of APWs,
while a reduction of $S_\gamma$ significantly accelerates the angular-momentum convergence.

\begin{figure}[b] 
\includegraphics[trim=1cm 1.0cm 0cm 1cm, clip=true,width=0.49\textwidth]{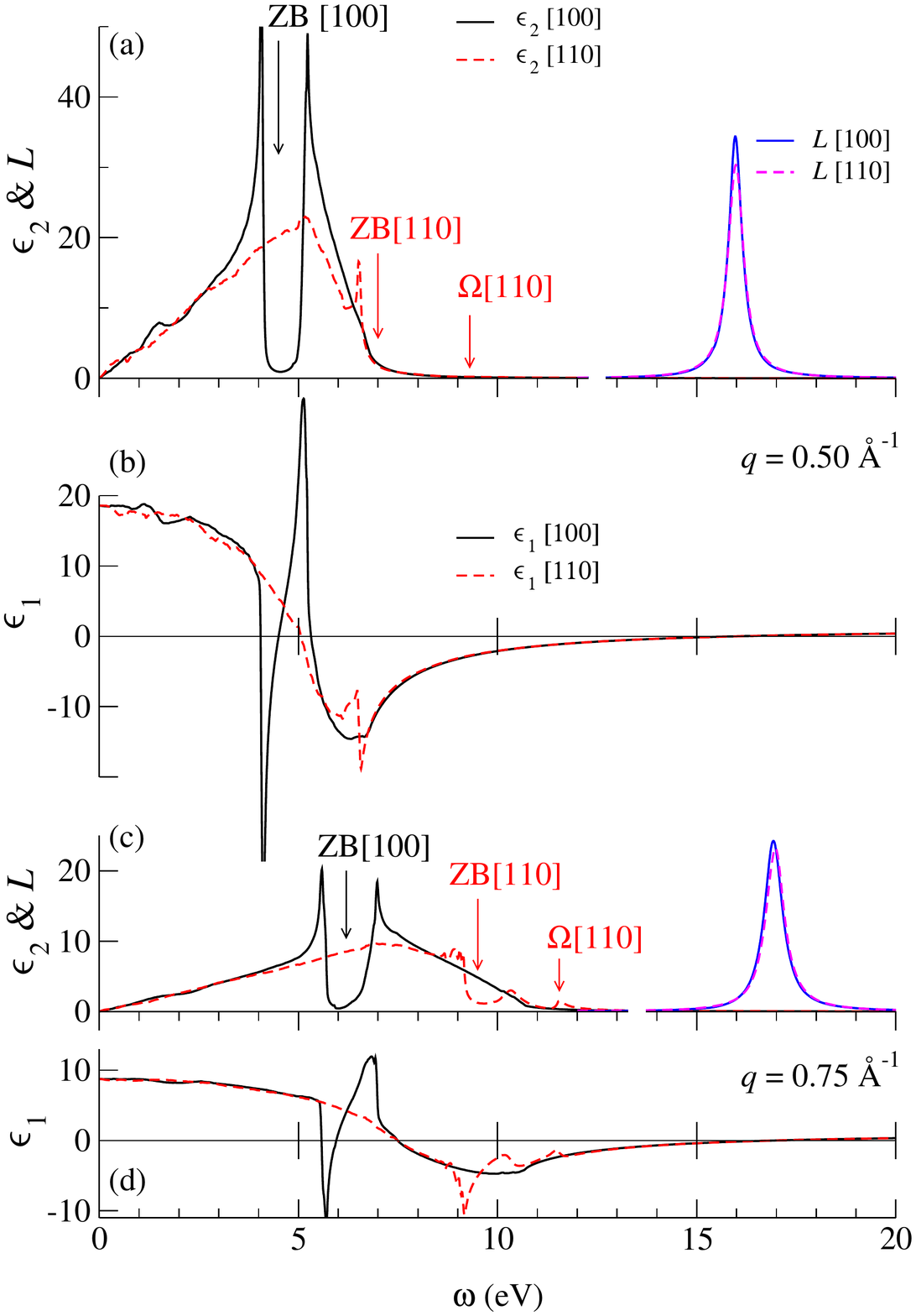}
\caption{Real $\epsilon_1$ and imaginary $\epsilon_2$ parts of the DF and loss function
  $L=-\text{Im}(1/\epsilon)$ of Al for $q=0.5$~\iaa, (a) and (b), and $0.75$~\iaa,
  (c) and (d). Full lines are for \qIOO\ and dashed lines for \qIIO. Vertical arrows marked
  ZB[110] indicate a zone-boundary gap, and $\Omega$[110] indicate a maximum at the
  high-energy slope of the spectrum, see Sec.~\ref{PD} and Fig.~\ref{fig-AN}.
}\label{fig-DF}
\end{figure}      
\section{Calculation of plasmon energy}\label{computation}
In this work we are mainly interested in the plasmon dispersion in Al and Na, where the plasmon
energy is well above the intense interband transitions and, at the same time, it is well below
the onset of semi-core excitations, 65~eV in Al and 25~eV in Na. In such cases the local
fields can be neglected to a good
approximation~\cite{Aryasetiawan_1994,Fleszar_1997,Cazzaniga_2011}, and DF reduces to the
following expression within the random phase approximation (RPA)~\cite{Ehrenreich_1959}:
\begin{equation}
\label{eq-cohendf}
\!\!\!\epsilon(\qv,\omega)=1-\frac{8\pi}{q^2V}\sum_{\lambda\lambda'\kv}
\frac{\left|\,\bra{\lambda'\kv\!+\!\qv}
\exp(i\qr)
\ket{\lambda\kv}\,\right|^2}{E_{\lambda'\kv+\qv}-
E_{\lambda\kv}-\hbar\omega+i\eta},
\end{equation}
where the summation is over occupied states $\ket{\lambda\kv}$ and unoccupied states
$\ket{\lambda'\kv+\qv}$. The imaginary part $\epsilon_2(\qv,\omega)$ is calculated in the limit
$\eta\to 0$ with the linear tetrahedron interpolation in $\kv$ space~\cite{LehmannTaut1972},
and the real part is obtained via the Kramers-Kronig relation. (The $\epsilon_2$ spectrum
cutoff was around 40~eV.) The convergence of the plasmon energies with respect to the $\kv$
point sampling is rather slow: For example, for Al, $\psi^{\kv}_{\lambda}$ are calculated on a
71$\times$71$\times$71 mesh in the reciprocal lattice cell, of which the irreducible
$\kv$-points are selected depending on the symmetry of the vector $\qv$. For \qIOO\ this
yields 47286 irreducible $\kv$-points and 273492 irreducible tetrahedra.

\begin{figure}[b] 
\includegraphics[trim=1cm 1cm 0cm 5cm, clip=true,width=0.49\textwidth]{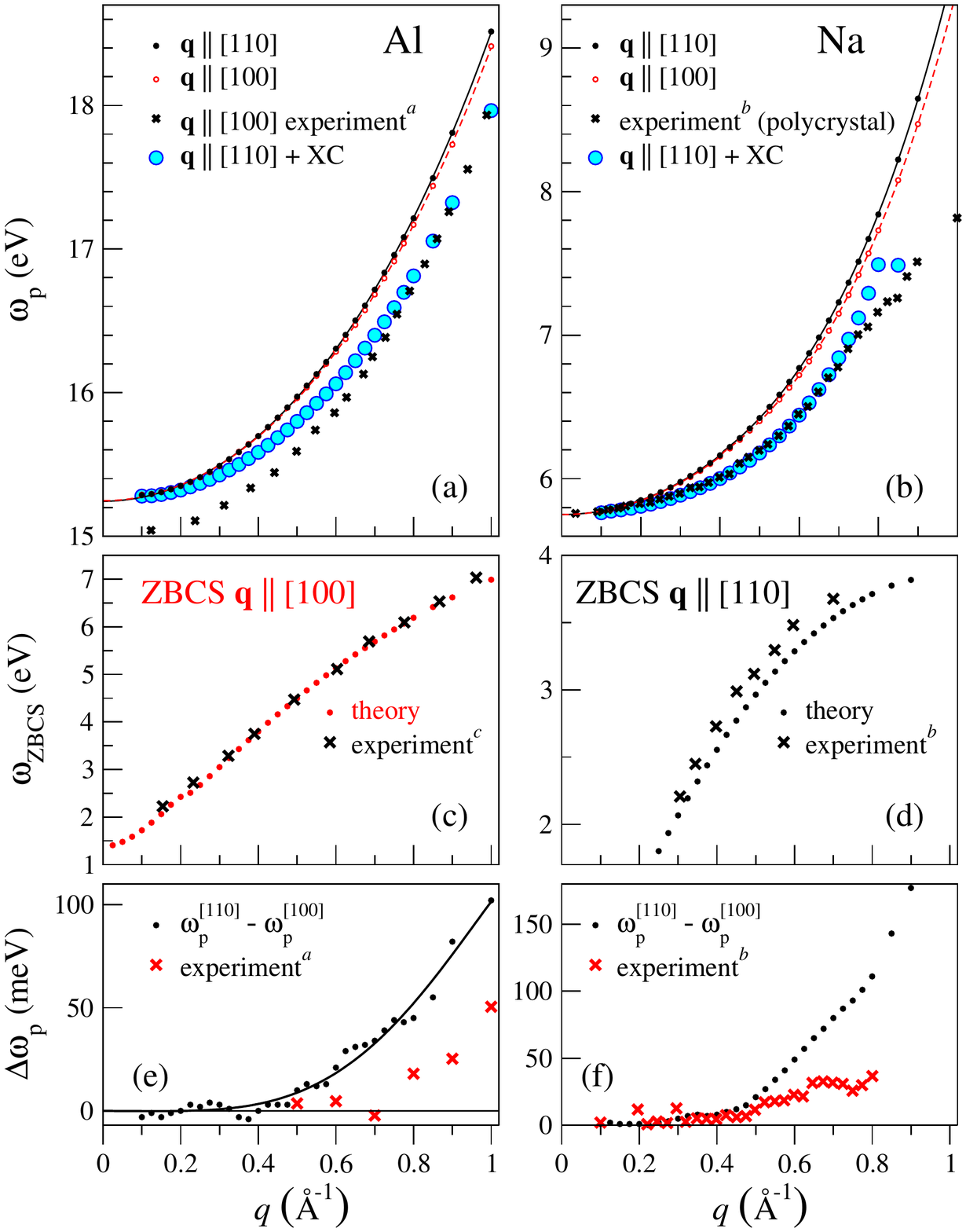}
\caption{Plasmon dispersion in Al (left column) and Na (right column). (a) and (b)
  Dispersion of the Drude plasmon for \qIIO\ (full black circles) and \qIOO\
  (open red circles) within the RPA and with the XC correction~\cite{Utsumi_1982}
  (large blue circles). (c) and (d) Dispersion of ZBCS for \qIOO\ for Al and for
  \qIIO\ for Na. (e) and (f) Anisotropy of the Drude plasmon. Crosses show the measurements
  of Refs.~\onlinecite{Sprosser_1989}$^a$, \onlinecite{vomFelde_1989}$^b$, and
  \onlinecite{Chen1977}$^c$ (extracted from graphical data).
}
\label{fig-PD}
\end{figure}

The complex DF of Al for \qIOO\ and \qIIO\ is shown in Fig.~\ref{fig-DF} for $q=0.5$ and
0.75~\iaa. The spectra agree well with the pseudopotential calculations by Lee and
Chang~\cite{Lee_1994}. Figure~\ref{fig-DF} also shows a high-energy part of the loss
function $L(\qv,\omega)=-\text{Im}[1/\epsilon(\qv,\omega)]$, whose maximum above the intense
interband transitions is referred to as the Drude plasmon because it can be related to the
jellium model~\cite{Lindhard1954}. The $\epsilon_2(\omega)$ spectrum is seen to be strongly
anisotropic: a zone-boundary gap (ZB) is very pronounced for \qIOO\ and is much weaker for
\qIIO. For \qIOO\ this structure gives rise to a very sharp low-energy peak in the loss
function, see Fig.~\ref{fig-PD}(c). This is the well-known zone boundary collective state
first predicted for simple metals by Foo and Hopfield~\cite{Foo_1968}.

The dispersion $\omega_{\rm p}(q)$ of the Drude plasmon is shown in Fig.~\ref{fig-PD}(a) for
\qIOO\ and \qIIO. Let us extrapolate the dispersion curves to $q=0$ by fitting the
calculated $\omega_{\rm p}(q)$ points with a Lindhard-like function
$\omega_{\rm p}(q)=\omega_{\rm p}(0)+\alpha q^2+\beta q^4$, see Table~\ref{tab}.
For Al we obtain $\omega_{\rm p}(0)=15.24$~eV, which is in accord with the value of
15.28~eV calculated in Ref.~\cite{Lee_1994} but strongly deviates from the measured value
$\omega_{\rm p}(0)=15.01\pm 0.01$~eV~\cite{Sprosser_1989}. For Na the agreement is much
better: $\omega_{\rm p}(0)=5.75$~eV in our theory and 5.76$\pm 0.02$~eV in the
experiment~\cite{vomFelde_1989} (extracted from graphical data).
\begin{table}[t]
\caption{\label{tab} Fitting the plasmon dispersion in Al with the function
$\omega_{\rm p}(q)=\omega_{\rm p}(0)+\alpha q^2+\beta q^4$. XC indicates that a
static exchange-correlation correction is included.}
\begin{ruledtabular}
\begin{tabular}{llll}
  & $\omega_{\rm p}(0)$ (eV) & $\alpha$ (eV\AA$^2) $ & $\beta$ (eV\AA$^4$) \\[1mm]
this work \qIOO\                & 15.24   & 2.75 &  0.39     \\
this work \qIIO\                & 15.24   & 2.77 &  0.49     \\
this work \qIOO\ (XC)           & 15.24   & 2.03 &  0.51     \\
this work \qIIO\ (XC)           & 15.24   & 2.04 &  0.65     \\
experiment \cite{Sprosser_1989} & 15.01   & 2.27 &  0.65     \\
theory~\cite{Lee_1994} (XC)     & 15.28   & 2.13 &  0.58
\end{tabular}
\end{ruledtabular}
\end{table}

Apart from fundamental approximations, the theoretical results suffer from the computational
uncertainty of $\epsilon(\omega)$. Indeed, the plasmon energy is determined by the condition
$\epsilon_1(\omega_{\rm p})=0$, and for a small slope $d\epsilon_1(\omega)/d\omega$ it becomes
very sensitive to small errors in $\epsilon_1(\omega)$. The most important source of error
is the inaccuracy of the wave functions $\braket{\rv}{\lambda\kv}$, which results in an error
in the numerator of Eq.~\eqref{eq-cohendf}. This is an inevitable shortcoming of the
variational wave functions, which stems from the incompleteness of the basis set. Its effect
on the accuracy of $\omega_{\rm p}$ can be estimated by comparing a numerical $q\to 0$ limit by
Eq.~\eqref{eq-cohendf} with the calculation in the optical limit $q=0$. In the latter case
the intraband part $\lambda=\lambda'$ acquires the Drude form
$\epsilon_1(\omega)=1-\omega_0^2/\omega^2$. For a cubic crystal, $\omega_0^2$ is the diagonal
element of the plasma frequency tensor $\omega_0=\omega_{xx}=\omega_{yy}=\omega_{zz}$ given by
the integral over the Fermi surface
\begin{equation}\label{eq-intraband}
\omega_{\mu\nu}^2=
\frac{1}{\pi^2}\sum_{\lambda}\int\frac{v_{\mu}(\lambda\kv)v_{\nu}(\lambda\kv)}
 {|\mathbf{v}(\lambda\kv)|}\,dS_{\rm F},
\end{equation}
where $\mu$ and $\nu$ indicate Cartesian components of the group velocity
$\mathbf{v}(\lambda\kv)$ in the state $\ket{\lambda\kv}$. For $\lambda\ne\lambda'$ the
numerator of Eq.~\eqref{eq-cohendf} reduces to the squared modulus of the momentum matrix
element
$\bra{\lambda'\kv}\mathbf{\hat{p}}\ket{\lambda\kv}$:
\begin{equation}\label{eq-limit}
\lim_{q\to0}
\bra{\lambda'\kv\!+\!\mathbf{q}}\!\exp(i\qr)\!\ket{\lambda\kv}=
\frac{\hbar}{m}
\frac{\bra{\lambda'\kv}\mathbf{\qv\!\cdot\!\hat{p}}\ket{\lambda\kv}}{E_{\lambda'\kv}-E_{\lambda\kv}}.
\end{equation}

In the optical limit we obtain for Al $\omega_{\rm p}^{\rm opt}=15.13$~eV and for Na
$\omega_{\rm p}^{\rm opt}=5.72$~eV, i.e., the uncertainty of $\omega_{\rm p}$ amounts to 0.11~eV in
Al and 0.03~eV in Na. According to our analysis, most of the error comes from the numerator of
the interband term in Eq.~\eqref{eq-cohendf}, and there are two reasons why in Al it is larger
than in Na. First, the slope of the $\epsilon_1(\omega)$ curve at $\omega_{\rm p}$ is much
smaller in Al than in Na: $d\epsilon_1(\omega)/d\omega=0.13$~eV$^{-1}$ in Al and 0.35~eV$^{-1}$
in Na. Second, at $q\to 0$ interband transitions play much larger role in Al than in Na. In Al,
the intrinsic uncertainty in $\epsilon_1(\omega_{\rm p})$ is 0.014, which is $\sim 4$\% of the
interband contribution. The error stems form the inaccuracy of the $\psi^{\kv}_{\lambda}$
themselves, and it far exceeds the error due to the approximate treatment of the products
$\psi^{\kv *}_{\lambda'}\psi^{\kv-\qv}_{\lambda}$ with a computationally reasonable product basis
set, as we show in Appendix.

\section{Anisotropy of plasmon dispersion}\label{PD}
The basic aspects of the DF of nearly-free-electron metals can be understood from the Lindhard
formula for jellium~\cite{Lindhard1954}, but the band structure effects have long been realized
to be important~\cite{Bross_1978}. In particular, zone boundary collective states were
identified in Al~\cite{Sturm_1984} and in Li and Na~\cite{Sturm_1989}. The relation between
the underlying band structure and the energy loss function was analyzed for
Al~\cite{Lee_1994,Maddocks1994,Kaltenborn_2013} and for alkali metals
\cite{TautSturm_1992,Quong_1993,Aryasetiawan_1994,Fleszar_1997,Ku_1999,Huotari2009}, also under
pressure~\cite{Silkin_2007,RodriguezPrieto_2008,Errea_2010,Loa_2011,Attarian_2014,Azpiroz_2014,Yu_2018}. In cesium, the interband transitions were
found to cause a negative plasmon dispersion \cite{Aryasetiawan_1994,Fleszar_1997}. Many works
addressed the role of the crystal local field and exchange and correlation (XC) in the
dielectric response: the effect of the many-body interactions on the plasmon dispersion was
studied for Al and for alkalis in Refs.~\onlinecite{Quong_1993,Cazzaniga_2011,Taut_1992,Aryasetiawan_1994,Karlsson_1995,Fleszar_1997,Ku_1999,Huotari2009,Nechaev_2007,RodriguezPrieto_2008,Azpiroz_2014}.
It was concluded that the local fields effects are almost negligible unless the semi-core
states are involved~\cite{Aryasetiawan_1994,Fleszar_1997,Cazzaniga_2011}.

Obviously, the anisotropic band structure should lead to the anisotropy of the loss function.
The directional dependence of the plasmon dispersion was studied experimentally in
Al~\cite{Urner_1976,Petri_1976,Sprosser_1989}, Na~\cite{vomFelde_1989}, and
Li~\cite{Schulke_1986} and theoretically analyzed using model
approaches~\cite{Sturm_1978,Sturm_1989,Bross_1978} and from first
principles~\cite{Taut_1992,Quong_1993,Lee_1994,Ku_1999,Nechaev_2007}. In Al and Na
the anisotropy of the Drude plasmon is rather subtle: experimentally, the difference
$\Delta\omega_{\rm p}=\omega_{\rm p}^{[110]}-\omega_{\rm p}^{[100]}$ reaches $\approx 0.05$~eV around
1~\iaa~\cite{Sprosser_1989,vomFelde_1989}, see Figs.~\ref{fig-PD}(e) and \ref{fig-PD}(f). The
theoretical estimates based on a local pseudopotential model of Bross~\cite{Bross_1978} and
on the {\it ab initio} pseudopotential study by Lee and Chang~\cite{Lee_1994} gave
$\Delta\omega_{\rm p}=0.2$ and 0.25--0.30~eV, respectively. On the contrary, in the
{\it ab initio} pseudopotential calculation by Quong and Eguiluz~\cite{Quong_1993}
no anisotropy was resolved.

\begin{figure}[t] 
\includegraphics[trim=11mm 12mm 00mm 180mm, clip=true,width=0.5\textwidth]{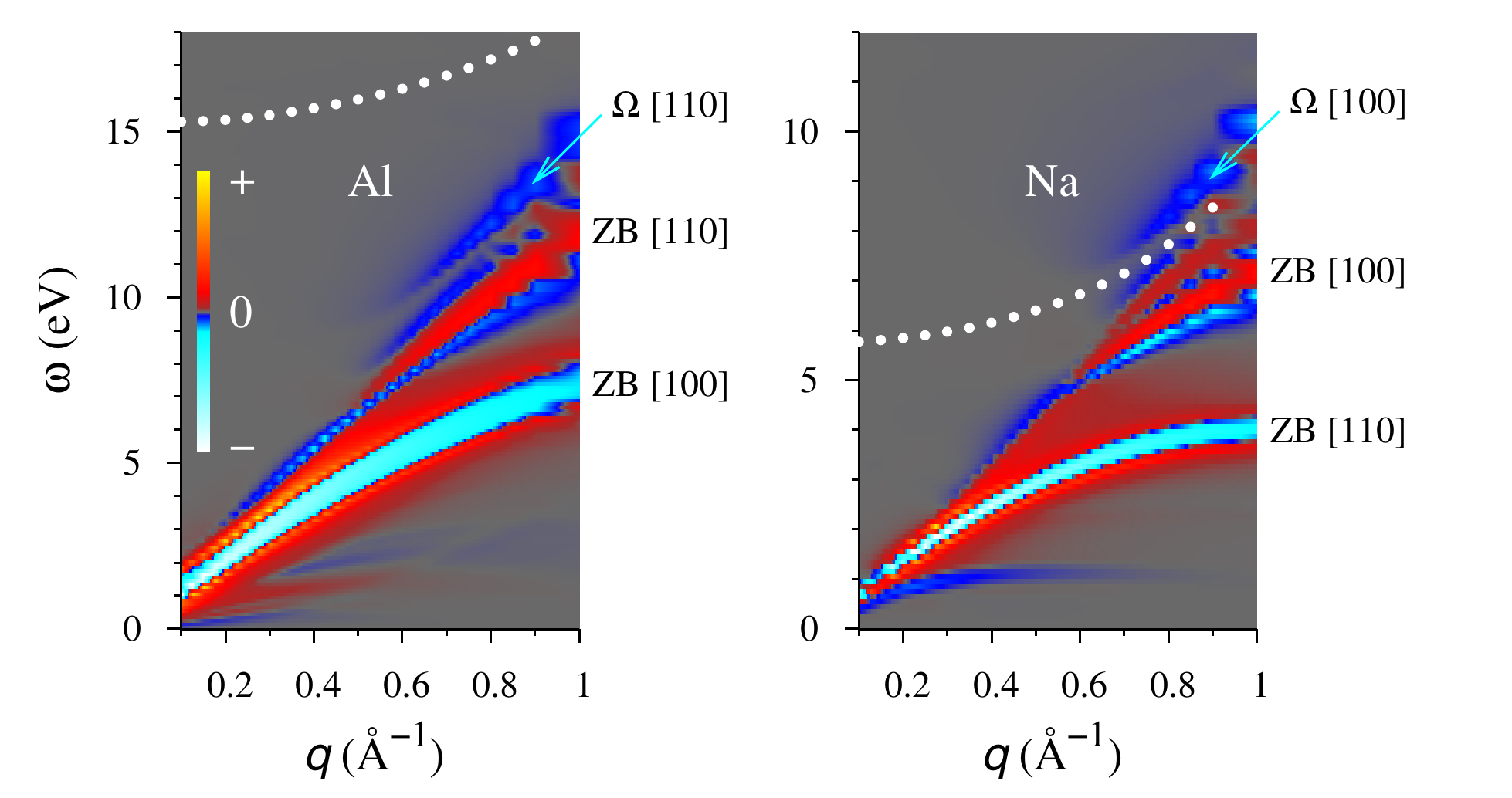}
\caption{Energy-momentum map of the anisotropy of $\epsilon_2$: left graph shows
  $\omega\epsilon_2^{[100]}-\omega\epsilon_2^{[110]}$ for Al and right graph shows
  $\omega\epsilon_2^{[110]}-\omega\epsilon_2^{[100]}$ for Na. White circles show the location of
  the Drude plasmon.}
\label{fig-AN}    
\end{figure}

It is tempting to relate the anisotropy of $\omega_{\rm p}(q)$ to the most conspicuous
anisotropic feature of the $\epsilon_2(\qv,\omega)$ spectrum, the zone-boundary gap, see
Fig.~\ref{fig-DF}. However, obviously, this feature cannot explain the observation. First,
both experimentally and theoretically, Al and Na behave oppositely with respect to the ZBCS:
in Al, ZBCS occurs for \qIOO\ and in Na for \qIIO, whereas both in Al and in Na it is
$\omega_{\rm p}^{[110]}>\omega_{\rm p}^{[100]}$. Second, $\Delta\omega_{\rm p}$ grows with
increasing $q$, whereas the influence of the ZB gap decreases, as illustrated by the
difference $\epsilon_2^{[100]}-\epsilon_2^{[110]}$ for Al (and opposite function for Na) shown in
Fig.~\ref{fig-AN}. The energy-momentum distributions of the anisotropy are seen to be very
similar in Al and Na, with Na[110] playing the role of Al[100] and Na[100] the role of Al[110].
In Al, the $\epsilon_2$ spectrum for \qIIO\ has a more complicated structure than for \qIOO:
apart from the low-energy gap (denoted ZB[110] in Figs.~\ref{fig-DF} and \ref{fig-AN}), at
larger $q$ there emerges a maximum (denoted $\Omega$[110]) at the high-energy slope of the
spectrum. Its contribution to $\epsilon_1(\omega_{\rm p})$ grows with $q$, and because it
appears below $\omega_{\rm p}^{[110]}$ it shifts $\omega_{\rm p}^{[110]}$ to higher energies. A
similar picture takes place for Na, only there is an additional dip-peak structure between
ZB[100] and $\Omega$[100], see Fig.~\ref{fig-AN}. However, the effect is the opposite: in
Na the $\Omega$[100] peak occurs {\em above} the plasmon energy, so that it shifts
$\omega_{\rm p}^{[100]}$ to lower energies.

Thus, the opposite behavior of the anisotropy of the Drude plasmon relative to ZBCS in Al and
Na is explained by the different location of the uppermost structure $\Omega$ relative to
$\omega_{\rm p}$. Furthermore, Fig.~\ref{fig-AN} explains why both in Al and in Na the anisotropy
is negligible below and rapidly grows above $q=0.5$~\iaa, see Figs.~\ref{fig-PD}(e) and
\ref{fig-PD}(f): it is at this wave vector that the $\Omega$-structure appears in the spectrum.

Within the RPA, the difference between the measured and the calculated
energies of the Drude plasmon is of the same order of magnitude in Al and Na. The discrepancy
is largely due to the neglect of exchange and correlation. These effects can be approximately
included by a static local field correction function $G(q)$ using the parametrization of Utsumi
and Ichimaru~\cite{Utsumi_1982}:
\begin{equation}\label{eq-LFXC}
\epsilon(\qv,\omega)\to
1+\frac{\epsilon(\qv,\omega)-1}{1-G(q)\left[\epsilon(\qv,\omega)-1\right]}.
\end{equation}
The XC correction brings the plasmon dispersion in Na into excellent agreement with the
experiment over the interval up to $q=0.75$~\iaa, see Fig.~\ref{fig-PD}(b). For Al, the XC
correction transforms the $\omega_{\rm p}(q)$ curve in a similar way, see Fig.~\ref{fig-PD}(a),
but the parameter $\alpha$ is strongly underestimated, see Table.

\begin{figure}[b] 
\includegraphics[trim=10mm 15mm 50mm 13mm, clip=true,width=0.5\textwidth]{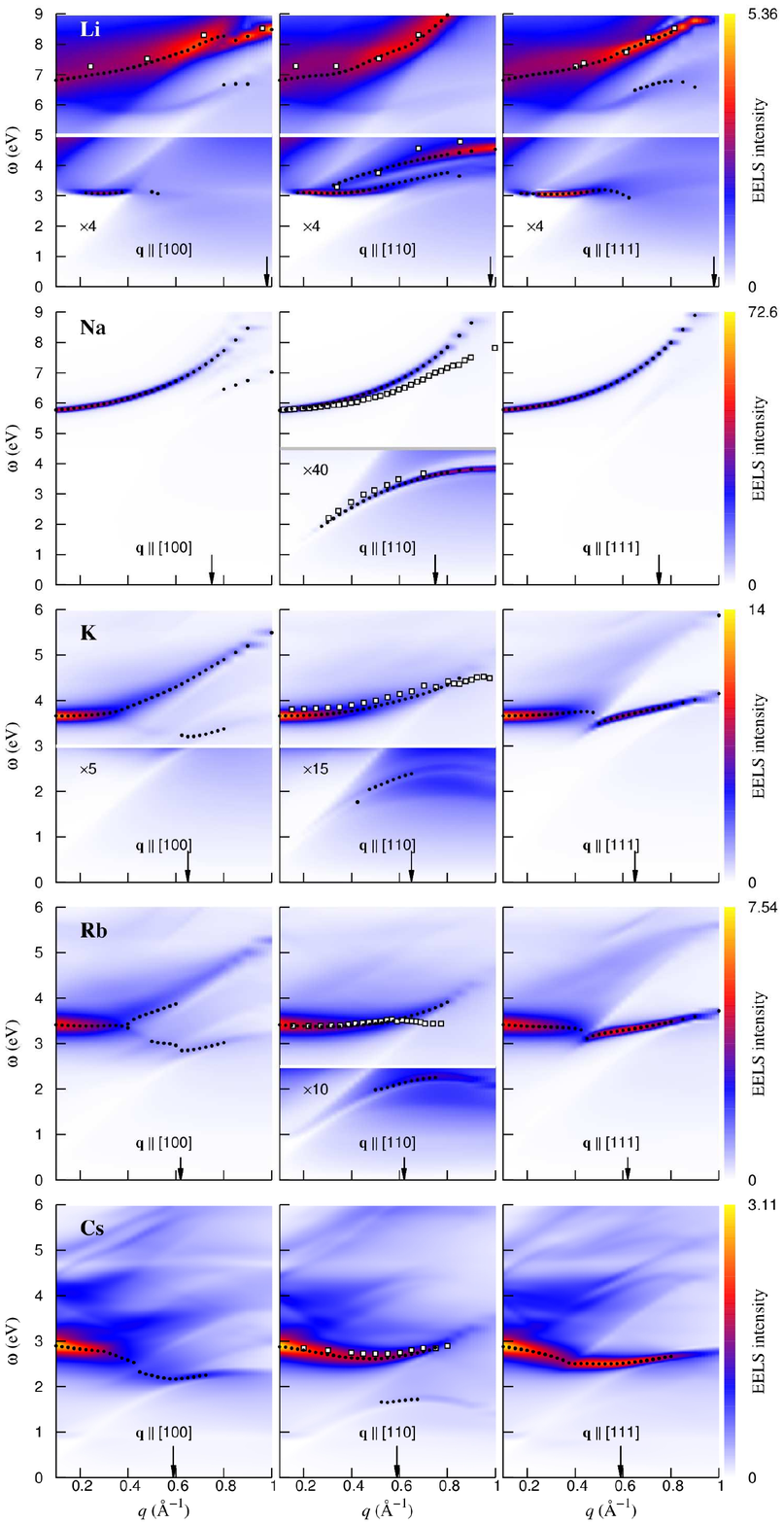}
\caption{Calculated energy loss spectra of Li, Na, K, Rb, and Cs for \qIOO, \qIIO, and \qIII.
White squares are  experimental data of Refs.~\onlinecite{Schulke_1986,Schulke_1984} for Li
and of Ref.~\onlinecite{vomFelde_1989} for Na, K, Rb, and Cs. Black circles show the plasmon
location, i.e., $\epsilon_1(q,\omega)=0$. The vertical arrows indicate the critical $q$
values of the Lindhard model. Note the scale change indicated by the horizontal white line.}
\label{fig-LiNaK}
\end{figure}

In contrast to Al and Na, in alkali metals Li, K, Rb, and Cs the interband transitions
around the plasmon energy are rather intense, see Fig.~\ref{fig-LiNaK}. Their strong energy
dependence causes more or less pronounced irregularities in the plasmon dispersion accompanied
by a strong damping (large spectral width) of the plasmon. Also the anisotropy of
$\epsilon_2(\omega)$ is very strong, which leads to a rather different shape of the
plasmon dispersion curve for different directions of $\qv$, see Fig.~\ref{fig-LiNaK}.

Figure~\ref{fig-LiNaK} show the loss function of alkali metals for \qIOO, \qIIO, and \qIII\
in comparison with the experiment. In Li, we again observe both the Drude plasmon at
$\omega_{\rm p}(0)=6.72$~eV and the ZBCS. For \qIIO, in our calculation two ZBCS branches are
resolved. In K, the plasmon peaks are much sharper than in Li, but the band-structure effects
are, evidently, very strong: the plasmon dispersion is far from parabolic, and its shape is
considerably different for different directions. For \qIIO, the agreement with the
experiment~\cite{vomFelde_1989} is rather good. (In contrast to Al and Na, in K the
experimental curve lies slightly above the theoretical one.) In Rb and Cs, the unoccupied
$3d$ states come closer to the Fermi level, and the strong dipole transitions to these states
are responsible for the pronouncedly non-free-electron behavior of the
DF~\cite{Aryasetiawan_1994}: in Rb the plasmon disperses only weakly, and in Cs the function
$\omega_{\rm p}(q)$ is non-monotonic, with a minimum at $\sim$0.5~\iaa\ in agreement with the
experiment~\cite{vomFelde_1989}. Our result is in good agreement with the pseudopotential
calculation of Ref.~\onlinecite{Fleszar_1997}.

\section{Summary}
We have developed a product basis set to represent response functions within an all-electron
framework. The proposed basis set consists of plane waves defined over the whole space and
spatially restricted orbitals, with the spatial dependence of the response function being
smoothly continuous by construction. It has a number of physically appealing and computationally
convenient properties. In particular, it can be defined universally, without regard to the
computational parameters of the band structure method (such as muffin-tin radii, energy
parameters, and number of APWs in LAPW). Consequently, the accuracy with which the physically
relevant response is represented can be set {\it a priori} and balanced with the intrinsic
accuracy of the underlying band structure calculation. Depending on the specific application,
the basis set can be optimized by tuning the relative weight carried by PWs and IOs.

We have implemented the scheme into the extended LAPW method and studied both the accuracy of
the representation of the dielectric function and the intrinsic accuracy of the underlying
wave functions.

We have applied the new method to the dielectric function of cubic $sp$ metals Al, Li, Na, K,
Rb, and Cs with the aim to understand what can be learned from the anisotropy of their bulk
loss function. In the non-free-electron-like metals Li, K, Rb, and Cs the anisotropy of the
plasmon dispersion is rather irregular due to the vicinity of the intra- and interband
transitions to the plasmon. In contrast, in Al and Na, the plasmon occurs well above the
intense interband transitions, and their effect is much tinier. We have reproduced the sign
and the qualitative trend of the $\qv$ dependence of the anisotropy in both metals, although
the absolute values are substantially overestimated in the calculation. We revealed a strong
similarity of the structure of the particle-hole transitions in Al and Na and traced the
plasmon anisotropy to the energy location of the plasmon relative to very subtle features of
the imaginary DF, which are barely visible in the $\epsilon_2(\omega)$ spectrum. Our analysis
demonstrates that the EELS experiment is sensitive to subtle details of the unoccupied bulk
band structure and may give access to features not reachable with angle-integrated (optics)
or surface-sensitive electron spectroscopies.
\begin{figure*}[t] 
\includegraphics[trim=0cm 12cm 0cm 0cm, clip=true,width=0.99\textwidth]{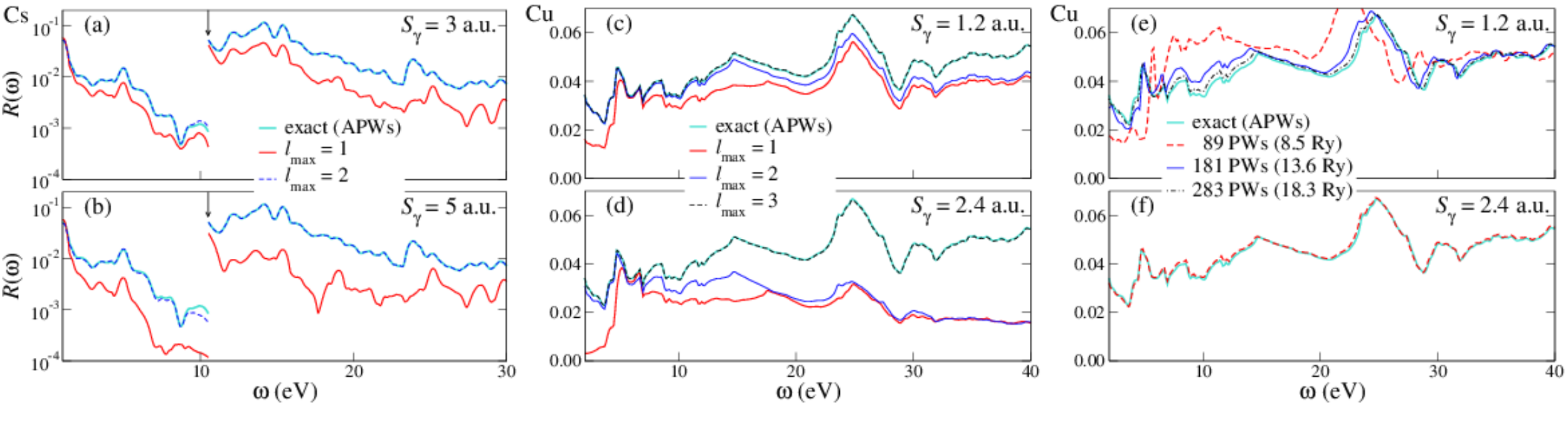}
\caption{Convergence of MME with the angular-momentum cutoff
  $l_\text{max}$, (a)--(d), and with the number of plane-waves, (e) and (f). For Cs,
  (a) and (b), $\kv$-averaged MME $R(\omega)$ are shown in logarithmic scale (a) for
  $S_\gamma=3$~a.u. and (b) for $S_\gamma=5$~a.u. The Fourier part comprises 229 PWs.
  Vertical arrows in the upper $\omega$-axes mark the onset of the semi-core transitions.
  For Cu, $S_\gamma=1.2$~a.u. in (c) and (e) and $S_\gamma=2.4$~a.u. in (d) and (f).
  In (c) and (d), the Fourier part comprises 459 PWs. In (e) and (f), the convergence
  with the number of PWs is for $l_{\text{max}}=4$. The cutoff energy $|\Gv|^2$ is given
  in parentheses.}
\label{fig-CONVOPT}
\end{figure*}      
\begin{figure*}[t] 
\includegraphics[trim=0cm 1cm 0cm 20cm, clip=true,width=0.99\textwidth]{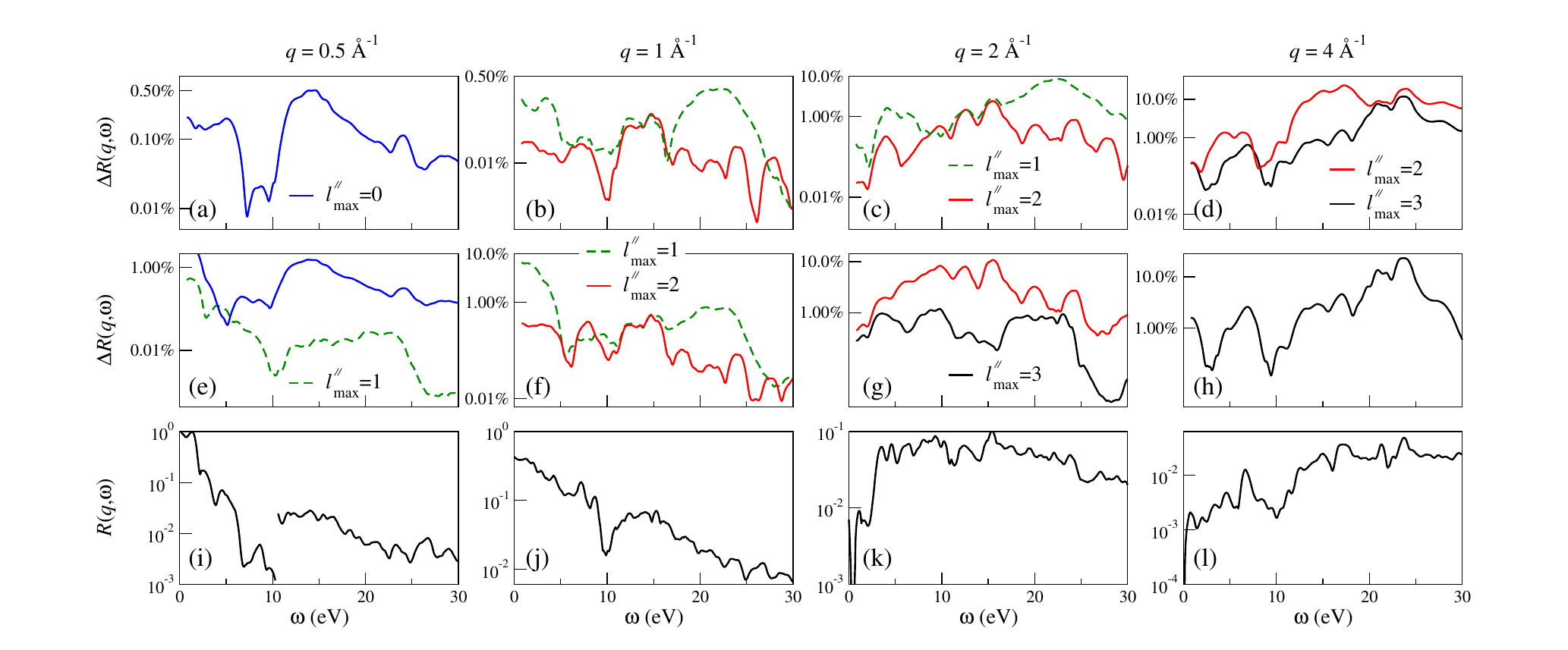}
\caption{Convergence of the Rayleigh expansion of $\exp(i\qr)$ for Cs for \qIOO\ for $q=0.5$,
  1.0, 2.0, and 4.0~\iaa for $S_\gamma=3$~a.u. (a)--(d) and for $S_\gamma=5$~a.u. (e)--(h).
  The lowest row shows the reference spectrum $\bar R(q,\omega)$ calculated with
  $l''_{\rm max}=4$, see Eq.~\eqref{eq-R}. The upper rows show $\kv$-averaged values of the
  error $\Delta R(l''_{\rm max})=|R(l''_{\rm max})-\bar R|$ with $l_{\rm max}=0$ to 3 [in percent
  relative to the average value of $\bar R(q,\omega)$ over the spectrum, note the logarithmic
  scale]. The Fourier part comprises 959 PWs.
}
\label{fig-convergCsFINITEQ}
\end{figure*}      

\begin{acknowledgments}
  We thank P.~Koval, R.~Kuzian, V.~Nazarov, I.~Nechaev, and V.~Silkin for enlightening
  discussions. This work was supported by the Spanish Ministry of Economy, Industry
  and Competitiveness MINEICO (Project No.~FIS2016-76617-P).
\end{acknowledgments}

\appendix*
\section{Accuracy and convergence}\label{accuracy}
Let us study the properties of the product basis set regarding the plane-wave cutoff of
$\psi^\text{F}$ and the angular-momentum cutoff of $\psi^\text{I}$, see Eq.~\eqref{eq-FI}.
We first consider the momentum operator $\hat\pv =-i\hbar\boldsymbol\nabla$, which is
related to the $q\to 0$ limit of the operator $\exp(i\QR)/q$, see Eq.~\eqref{eq-limit}.
This will give the idea of the accuracy of the filtered wave functions because the operator
$\hat\pv$ itself is treated exactly, and the result can be compared with the exact value
calculated from the complete APW representation of the wave function. To have an idea of
the accuracy of the momentum matrix elements (MME), let us consider its $\kv$-averaged value,
which we define as the ratio of the absorption probability
\begin{equation}
\label{rpa_nolf}
W(\omega) = \sum_{mn}\,\int_{\rm BZ}\!\!\!d\kv\,P^\kv_{mn}\,\delta(E_n^\kv-E_m^\kv-\hbar\omega)
\end{equation}
to the joint density of states $J(\omega)$, which is obtained by setting the
transition probability $P^{\mathbf{k}}_{mn}$ in Eq.~(\ref{rpa_nolf}) to unity,
\begin{equation}\label{eq-eps2}
R(\omega)=W(\omega)/J(\omega).
\end{equation}
The sum in Eq.~(\ref{rpa_nolf}) runs over all occupied states $\ket{n\kv}$ and unoccupied
states $\ket{m\kv}$. For a cubic crystal the dipole transition probability is
$P^{\mathbf{k}}_{mn} = \frac{1}{3}\left|\bra{m\kv}\hat\pv\ket{n\kv}\right|^2$.

It follows from the dipole selection rules that for reasonably small values of $S_\gamma$
the $l_\text{max}$-cutoff in Eq.~(\ref{eq-R}) does not need to exceed the highest
angular momentum of the atomic valence shell plus one. This is illustrated for Cs in
Figs.~\ref{fig-CONVOPT}(a) and \ref{fig-CONVOPT}(b) and for Cu in
Figs.~\ref{fig-CONVOPT}(c) and \ref{fig-CONVOPT}(d). Surprisingly, for Cs, not only
for a moderate $S_\gamma=3$~a.u. but also for a very large $S_\gamma=5$~a.u. at $l_\text{max}=2$
the error is negligible. Note that the results with the smaller $S_{\gamma}$ are slightly more
accurate although a larger part of the wave function is described by plane waves.
While for Cs the challenging aspect is the angular-momentum expansion of $\psi^\text{I}$ in a
large sphere, for Cu it is the plane-wave expansion of $\psi^\text{F}$.
Figures~\ref{fig-CONVOPT}(e) and ~\ref{fig-CONVOPT}(f) show the  PW-convergence of
the momentum matrix elements (MME) for $S_{\gamma}=1.2$ and 2.4~a.u., respectively.
Note that for $S_{\gamma}=2.4$~a.u., which is close to the MT radius, $R(\omega)$
converges already at 89~PWs, which equals the number of APWs needed to obtain the
band structure. The obvious advantage of the new basis is
that PWs are orthogonal and the momentum operator is diagonal.

For the operator $\exp(i\qr)$ there arises the question of the convergence of its Rayleigh
expansion, i.e., of the sum over $l''$ in Eq.~\eqref{eq-R}. The convergence of the
$\kv$-averaged matrix element $\left|\bra{m\kv+\qv}\exp(i\qr)\ket{n\kv}\right|^2$ is
illustrated in Fig.~\ref{fig-convergCsFINITEQ} for Cs for $S_\gamma=3$ and 5~a.u. The role of
quadrupole transitions is seen to increase with $q$, especially for the larger $S_\gamma$.
Interestingly, the quadrupole transitions from the semicore $5p$ states are more intense than
from the valence band. Figure~\ref{fig-convergCsFINITEQ} demonstrates that the $l''$ convergence
can be accelerated by diminishing the $\gamma$-sphere.

%


\end{document}